\documentclass[runningheads]{llncs}
\usepackage{amsmath}
\usepackage{mathdots,yhmath}
\usepackage{cancel,wasysym}
\usepackage{color}
\usepackage{array,multirow}
\usepackage{amssymb,gensymb}
\usepackage{appendix}
\usepackage{tabularx,booktabs}
\usepackage{graphicx}
\usepackage{svg}
\newcommand*{\Det}{\mbox{\textsf{Det}}}
\newcommand*{\true}{\mbox{\textsf{T}}}
\newcommand*{\false}{\mbox{\textsf{F}}}
\usepackage{footnote}
\makesavenoteenv{tabular}
\makesavenoteenv{table}

\usepackage{epstopdf}
\begin{document}
\title{Keyed Non-Parametric Hypothesis Tests}
\subtitle{Protecting Machine Learning From Poisoning Attacks}

\titlerunning{Keyed Hypothesis Tests}

\author{Yao Cheng\inst{1} \and
Cheng-Kang Chu\inst{1} \and
Hsiao-Ying Lin\inst{1}\and\\ 
Marius Lombard-Platet\inst{2,3} \and
David Naccache\inst{2}}
\authorrunning{Yao Cheng et al.}

\institute{Huawei International, Singapore\\
\email{\{chengyao101, chu.cheng.kang, lin.hsiao.ying\}@huawei.com} \and
DIENS, \'Ecole normale supérieure, CNRS, PSL Research University, Paris, France\\
\email{\{marius.lombard-platet, david.naccache\}@ens.fr}\and
Be-studys, Geneva, Switzerland
}

\maketitle 
\setcounter{footnote}{0}  

\begin{abstract}
The recent popularity of machine learning calls for a deeper understanding of AI security. Amongst the numerous AI threats published so far, poisoning attacks currently attract considerable attention.\smallskip

In a poisoning attack the opponent partially tampers the dataset used for learning to mislead the classifier during the testing phase.\smallskip

This paper proposes a new protection strategy against poisoning attacks.\smallskip

The technique relies on a new primitive called \textsl{keyed non-parametric hypothesis tests} allowing to evaluate \textsl{under adversarial conditions} the training input's conformance with a previously learned distribution $\mathfrak{D}$. To do so we use a secret key $\kappa$ unknown to the opponent.\smallskip

Keyed non-parametric hypothesis tests differs from  classical tests in that the secrecy of $\kappa$ prevents the opponent from misleading the keyed test into concluding that a (significantly) tampered dataset belongs to $\mathfrak{D}$. 

\keywords{Poisoning \and Machine learning security \and Hypothesis tests}
\end{abstract}

\section{Introduction \& Formalism}

The recent popularity of machine learning calls for a deeper understanding of AI security. Amongst the numerous AI threats published so far, poisoning attacks currently attract considerable attention.\smallskip

An ML algorithm $\mathcal{A}$ is a state machine with a two-phase life-cycle: during the first phase, called \textsl{training},  $\mathcal{A}$ builds a \textsl{model} (captured by a state variable $\sigma_i$) based on sample data $D$, called ``training data'': 

$$D=\{d_1,\ldots,d_k\} \mbox{~where~} d_i=\{\mbox{data}_i,\mbox{label}_i\}$$ 

Learning is hence defined by:

$$\sigma_i\leftarrow\mathcal{A}(\mbox{learn},\sigma_{i-1},d_i)$$

e.g. $\mbox{data}_i$ can be human face images and the  $\mbox{label}_i\in\{\mars,\female\}$.

During the second phase, called \textsl{testing}\footnote{Or \textsl{inference}.}, $\mathcal{A}$ is given an unlabelled $\mbox{data}$. $\mathcal{A}$'s goal is to predict as accurately as possible the corresponding $\mbox{label}$ given the distribution $\mathfrak{D}$ inferred from $D$.

$$\underline{\mbox{label}}=\mathcal{A}(\mbox{test},\sigma_{k},\mbox{data})$$

We denote by $T$ the dataset  $\{\mbox{data}_i,\mbox{label}_i\}$ used during testing where $\mbox{label}_i$ is the correct label (solution) corresponding to $\mbox{data}_i$ and $\underline{\mbox{label}}_i$ is the label predicted by $\mathcal{A}(\mbox{test},\sigma_{k},\mbox{data}_i)$\footnote{i.e. if $\mathcal{A}$ is perfect then \underline{\mbox{label}}=\mbox{label}.}.\smallskip

In a poisoning attack the opponent partially tampers $D$ to influence $\sigma_k$ and mislead $\mathcal{A}$ during testing. Formally, letting $\overline{d}=\{\overline{\mbox{data}},\overline{\mbox{label}}\}$, the attacker generates a \textsl{poison} dataset
$$\Tilde{D}=\{\Tilde{d}_1,\ldots,\Tilde{d}_k\}$$

resulting in a corrupted model $\Tilde{\sigma}_k$ such that 

$$\overline{\mbox{label}}\neq\underline{\overline{\mbox{label}}}=\mathcal{A}(\mbox{test},\Tilde{\sigma}_k,\overline{\mbox{data}})$$

Poisoning attacks were successfully implemented by tampering both incremental and periodic training models. In the \textsl{incremental training model} \footnote{Also called the \textsl{incremental update model}.}, whenever a new $d_i$ is seen during testing, $\mathcal{A}$'s performance on $d_i$ is evaluated and $\sigma$ is updated. In the \textsl{periodic retraining model}, data is stored in a buffer. When $\mathcal{A}$ falls below a performance threshold (or after a fixed number of queries) the buffer's data is used to retrain $\mathcal{A}$ anew. Retraining is either done using the buffer alone (resulting in a totally new $\sigma$) or by merging the buffer with previous information (updating $\sigma$).\smallskip

Protections against poisoning attacks can be categorized into two types: \textsl{robustification} and \textsl{sanitizing}: 

\subsubsection{Robustification} (built-in resistance) modifies $\mathcal{A}$ so that it takes into account the poison but tolerates its effect. Note that $\mathcal{A}$ does not need to identify the poisoned data as such but the effect of poisonous data must be diminished, dampened or nullified up to a point fit for purpose.\smallskip

The two main robustification techniques discussed in the literature are:\smallskip

\textsl{Feature squeezing} \cite{XEQ17,SGY+18} is a model hardening technique that reduces data complexity so that adversarial perturbations disappear because of low sensitivity. Usually the quality of the incoming data is degraded by encoding colors with fewer values or by using a smoothing filter over the images. This maps several inputs onto one ``characteristic'' or ``canonical input'' and reduces the perturbations introduced by the attacker. While useful in practice, those techniques inevitably degrade the $\mathcal{A}$'s accuracy.\smallskip

\textsl{Defense-GANs} \cite{SKC18} use Generative Adversarial Networks \cite{GPM+14} to reduce the poison's efficiency.  Informally, the GAN builds a model of the learned data and projects the input onto it.\smallskip

\subsubsection{Sanitizing} detects (by various methods e.g. \cite{CSL+08,LP16}) and discards poisoned $d_i$s. Note that sanitizing necessarily decreases $\mathcal{A}$'s ability to learn. \smallskip

This work prevents poisoning by sanitizing.\smallskip

\begin{figure}
\centering
\includegraphics[width=0.5\textwidth]{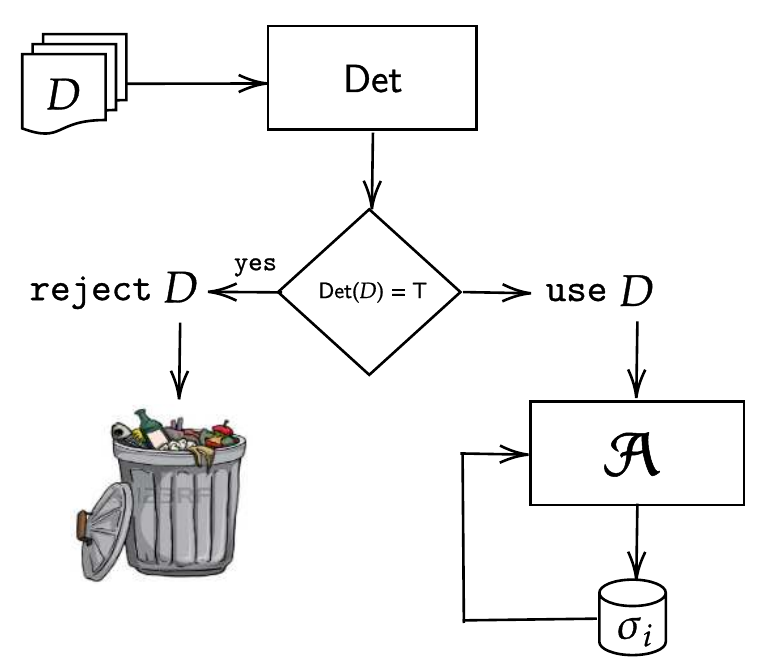}
\caption{Before entering $\mathcal{A}$, $D$ is given to the poison detection module $\Det$. If $\Det$ decides that $D$ is poisoned, $D$ is trashed. Otherwise $D$ is fed into $\mathcal{A}$ who updates $\sigma$.} \label{learning}
\end{figure}

Figure \ref{learning} shows a generic abstraction of sanitizing. $\mathcal{A}$ takes $D$ (periodically or incrementally) and outputs a $\sigma$ for the testing phase. But $d_i$s go through the poisoning detection module $\Det$ before entering $\mathcal{A}$. If $\Det$ decides that the probability that some $d_i$ is poisoned is too high, the suspicious $d_i$ is trashed to avoid corrupting $\sigma$.\smallskip

Because under normal circumstances $D$ and $T$ are drawn from the same distribution $\mathfrak{D}$ it is natural to implement $\Det$ using standard algorithms allowing to test the hypothesis $D\in\mathfrak{D}$.\smallskip

The most natural tool allowing one to do so is \textsl{nonparametric hypothesis tests} (NPHTs, hereafter denoted by $G$). Let $A,B$ be two datasets. $G(A,B)\in\{\true,\false\}$ allows to judge how compatible is a difference observed between $A$ and $B$ with the hypothesis that $A,B$ were drawn from the same distribution $\mathfrak{D}$.\smallskip

It is important to underline that $G$ is nonparametric, i.e. $G$ makes no assumptions on $\mathfrak{D}$.\smallskip

The above makes NPHTs natural candidates for detecting poison. However, whilst NPHTs are very good for natural hypothesis testing, they succumb spectacularly in adversarial scenarios where the attacker has full knowledge of the target's specification \cite{Kerk}. Indeed, section \ref{whatever} illustrates such a collapse.\smallskip

To regain a head-up over the attacker, our strategy will consist in mapping $A$ and $B$ into a \textsl{secret} space unpredictable by the adversary where $G$ can work confidentially. This mapping is defined by a key $\kappa$ making it hard for the adversary to design $A\in\mathfrak{D}$ and $B\in\mathfrak{D}'$ such that 

$$G(A,B)=\true\mbox{~~and~~}\mathfrak{D}\neq\mathfrak{D}'$$

\section{A Brief Overview of Poisoning Attacks}

Barreno et al. \cite{BNJT10} were the first to coin the term ``poisoning attacks''. Follow-up works such as Kearns et al. \cite{KL93} sophisticated and theorized this approach.\smallskip

Classical references introducing poisoning are \cite{PDL+06,NKS06,NBC+08,RNH+09,BNL12,XXE12,NPXN14,XBB+15,MZ15,BL17,KL17}. At times (e.g. \cite{KL17}) the opponent does not create or modify $d_i$'s but rather adds legitimate but carefully chosen $d_i$'s to $D$ to bias learning. Those inputs are usually computed using gradient descent. This was later generalized by \cite{SKC18}.\smallskip

During a poisoning attack, data modifications can either concern data or labels. \cite{BNL11} showed that a random flip of 40\% of labels suffices to seriously affect SVMs. \cite{KL10} showed that inserting malicious points into $D$ could gradually shift the decision boundary of an anomaly detection classifier. Poisoning points were obtained by solving a linear programming problem maximizing the mean of the displacement of the mass center of $D$. For a more complete overview we recommend \cite{BR18}.\smallskip

\subsubsection{Adversarial Goals.} Poisoning may seek to influence the classifier's decision when presented with a later target query or to leak information about $D$ or $\sigma$.\smallskip

The attacker's goals always apply to the testing phase and may be:

\begin{itemize}
    \item \textsl{Confidence Reduction}: Have $\mathcal{A}$ make more errors. In many cases, ``less confidence'' can clear suspicious instances at the benefit of doubt (\textsl{in dubio pro reo}).
    \item \textsl{Mis-classification attacks}: are defined by replacing $\mbox{\textsf{adj}}_1,\mbox{\textsf{adj}}_2$ in the definition:\smallskip
\begin{center}
    ``Make  $\mathcal{A}$ conclude that a $\mbox{\textsf{adj}}_1$ $\mbox{data}_i$ belongs to a $\mbox{\textsf{adj}}_2$ wrong label.''    
\end{center}    
   \end{itemize}
\begin{center}

\begin{tabular}{|l|c|c|}\hline
\textbf{~~Attack}                      & $\mbox{\textsf{adj}}_1$ & $\mbox{\textsf{adj}}_2$  \\\hline
~~Mis-classification\footnote{This is useful if any mistake may serve the opponent's interests e.g. any navigation error would crash a drone with high probability.}                   &~~random~~&~~random~~\\\hline
~~Targeted Mis-classification          &~~chosen~~&~~random~~\\\hline
~~Source-Target Mis-classification~~&~~chosen~~&~~chosen~~\\\hline
\end{tabular}
\end{center}

 \subsubsection{Adversarial capabilities} designate the degree of knowledge that the attacker has on the target system. Bibliography distinguishes between \textsl{training phase capabilities} and \textsl{testing phase capabilities}. Poisoning assumes training phase capabilities.\smallskip 

The attacker's capabilities may be:

\begin{itemize}
    \item \textsl{Data Injection}: Add new data to $D$.
    \item \textsl{Data Modification}: Modify $D$ before training.
    \item \textsl{Logic Corruption}: Modify the code (behavior) of $\mathcal{A}$\footnote{This is the equivalent of fault attacks in cryptography.}.
\end{itemize}

\section{Keyed Anti-Poisoning}

To illustrate our strategy, we use Mann-Whitney's $U$-test and Stouffer's method that we recall in the appendix.\smallskip

We assume that when training starts, we are given a safe subset of $D$ denoted $D_s$ (where the subscript $s$ stands for ``safe''). Our goal is to assess the safety of the upcoming subset of $D$ denoted $D_u$ (where the subscript $u$ stands for ``unknown'').

We assume that $D_s$ and $T$ come from the same distribution $\mathfrak{D}$. As mentioned before, the idea is to map $D_s$ and $D_u$ to a space $f_{\kappa}(\mathfrak{D})$ hidden from the opponent. $f$ is keyed to prevent the attacker from predicting how to create adversarial input fooling $\mathcal{A}$.\smallskip

\begin{figure}
\centering
\includegraphics[width=0.5\textwidth]{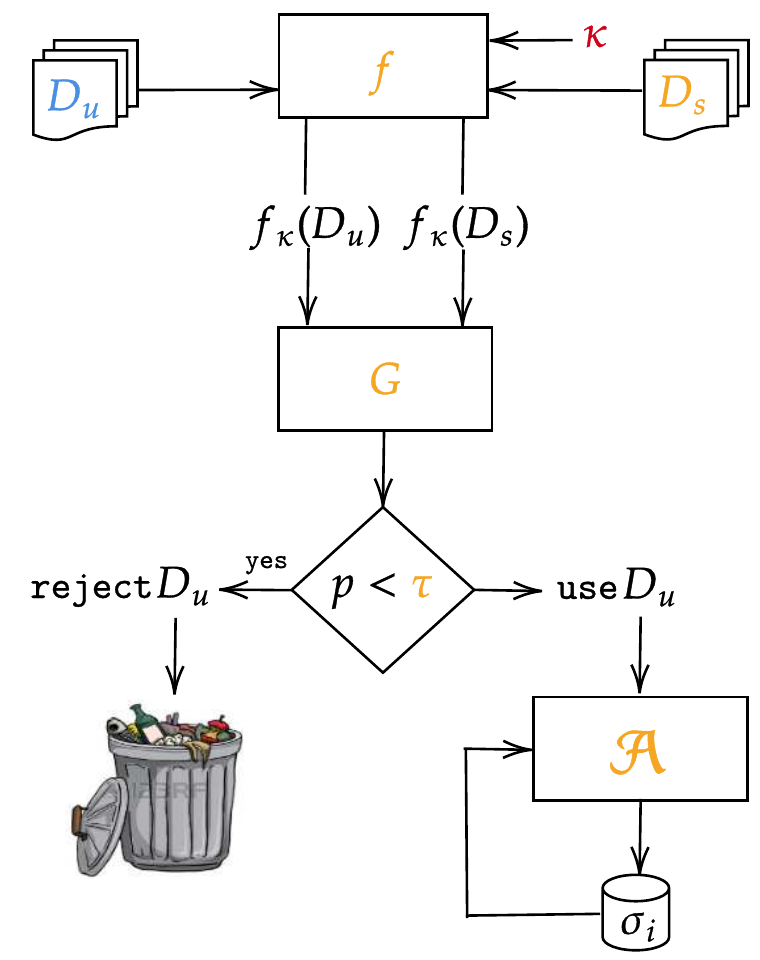}
\caption{Implementing \Det~using keying. \textcolor{red}{Red}: unknown to the opponent. \textcolor{orange}{Orange}: known by the opponent.
\textcolor{blue}{Blue}: controlled by the opponent.} \label{keyed}
\end{figure}

Figure \ref{keyed} shows the details of the $\Det$ plugin added to $\mathcal{A}$ in Figure \ref{learning}. $\Det$ takes a key $\kappa$, reads $D_s,D_u$, performs the keyed transformation, calls $G$ on $f_{\kappa}(D_u),f_{\kappa}(D_s)$ and outputs a decision.\smallskip 

$G$ can be Mann-Whitney's test (illustrated in this paper) or any other NPHT e.g. the location test, the paired $T$ test, Siegel-Turkey's test, the variance test, or multidimensional tests such as deep gaussianization \cite{Tolpin:2019:PAD:3297280.3297414}.\smallskip

\subsection{Trivial Mann-Whitney Poisoning}
\label{whatever}

Let $G$ be Mann-Whitney's $U$-Test returning a $p$-value:

\[
p = G(A,B) = G(\{a_0,\ldots,a_{n_1-1}\}, \{b_0,\ldots,b_{n_2-1}\})
\]

\begin{figure}
\centering
\includegraphics[width=0.7\textwidth]{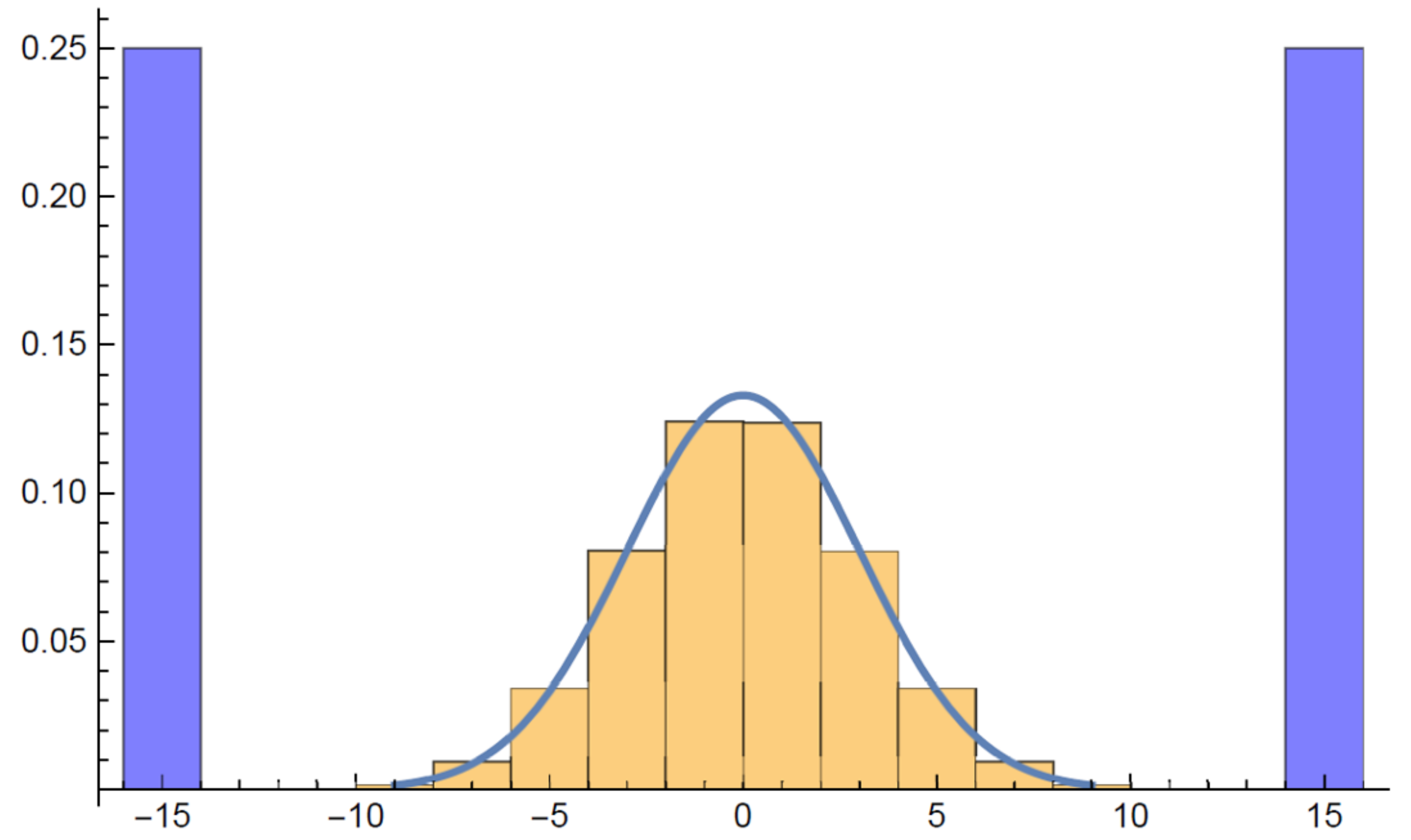}
\caption{Trivial Mann-Whitney poisoning. Samples drawn from the blue distribution are Mann-Whitney-indistinguishable from samples drawn from the orange one.} \label{attackMW}
\end{figure}

$G$ is, between others, susceptible to poisoning as follows: assume that $A$ is sampled from a Gaussian distribution $A\in_R\mathcal{N}(\mu_A,\sigma_A)$ and that $B$ is sampled from $\{-q,q\}$ where $\mu_A\ll q$ (Figure \ref{attackMW}). While $A$ and $B$ are totally different, $G$ will be misled.\smallskip

For instance, after picking $10^6$ samples $A \in_R \mathcal{N}(0,3)$ and $10^6$ samples $B \in_R \{-15,15\}$ (i.e. we took $q=15$), we get a $p$-value of $0.99$. From Mann-Whitney's perspective, $A,B$ come from the same parent distribution with a very high degree of confidence while, in all evidence, they do not.\smallskip

\subsection{Keyed Mann-Whitney}
We instantiate $f_{\kappa}$ by secret random polynomials i.e. polynomials $R(x)$ whose coefficients are randomly and secretly refreshed before each invocation of $G$. Instead of returning $G(A,B)$, $\Det$ returns $G(R(A),R(B))$ where:
\[
R(A) = \Big\{R(a_0),\dots,R(a_{n_1-1})\Big\} \quad\mbox{and}\quad R(B) = \Big\{R(b_0),\dots,R(b_{n_2-1})\Big\}
\]

The rationale is that $R$ will map the attacker's input to an unpredictable location in which the Mann-Whitney is very likely to be safe.\smallskip

$\ell$ random polynomials $R_1(x),\dots,R_{\ell}(x)$ are selected as keys and $\Det$ calls $G$ for each polynomial. To aggregate all resulting $p$-values, $\Det$ computes:

\[
\Delta = \mbox{Stouffer}\Big(G\Big(R_1(A), R_1(B)\Big),\ldots,G\Big(R_{\ell}(A), R_{\ell}(B)\Big)\Big)
\]

If $\Delta \approx 0$, the sample is rejected as poisonous with very high probability.\smallskip

Note that any smooth function can be used as $R$, e.g. B-splines. The criterion on $R$ is that the random selection process must yield significantly different functions.

\subsection{Experiments}
We illustrate the above by protecting $\mathcal{N}(0,1)$. The good thing about $\mathcal{N}(0,1)$ is that random polynomials tend to diverge when $x=1$ but adapt well to the central interval in which the Gaussian is not negligible.\smallskip

We attack $\mathcal{N}(0,1)$ by poisoning with $\{-q, q\}$, where $q$ is set to 3, 2, 1, and 0.5, respectively. For each value of $q$, two sets of 50 samples are drawn from the two distributions. Those samples are then transformed into other sets by applying a random polynomial of degree 4 and then fed into $G$ to obtain a $p$-value (using the two-sided mode). This $p$-value predicts whether these two sets of transformed samples come from the same distribution: a $p$-value close to 0 is a strong evidence against the null hypothesis. In each of our experiments, we apply nine secret random polynomials of degree 4 and aggregate the resulting $p$-values using Stouffer's method. For each setting, we run 1000  simulations. Similarly, for the same polynomials and $q$, we run a ``honest'' test, where both samples come from the same distribution.\smallskip

We thus retrieve 1000 ``attack'' $p$-values, which we sort by ascending order. Similarly, we sort the ``honest'' $p$-values. It is a classic result that, under the null hypothesis, a $p$-value follows a random uniform distribution over $[0, 1]$, hence a plot of the sorted ``honest'' $p$-values is a linear curve.\smallskip

An attack is successful if, on average, the ``attack'' sample is accepted as least as often as the ``honest'' sample. This can be rewritten as $\mathrm E(p^{\mbox{\scriptsize{attack}}}) \geq \mathrm E(p^{\mbox{\scriptsize{honest}}})$, with $E$ the . Hence, a sufficient condition for the validity is that the curve of sorted attack $p$-values (solid lines in our figures) is above the curve of sorted honest $p$-values (dashed lines). \smallskip

\begin{figure}
\includegraphics[width=0.9\textwidth]{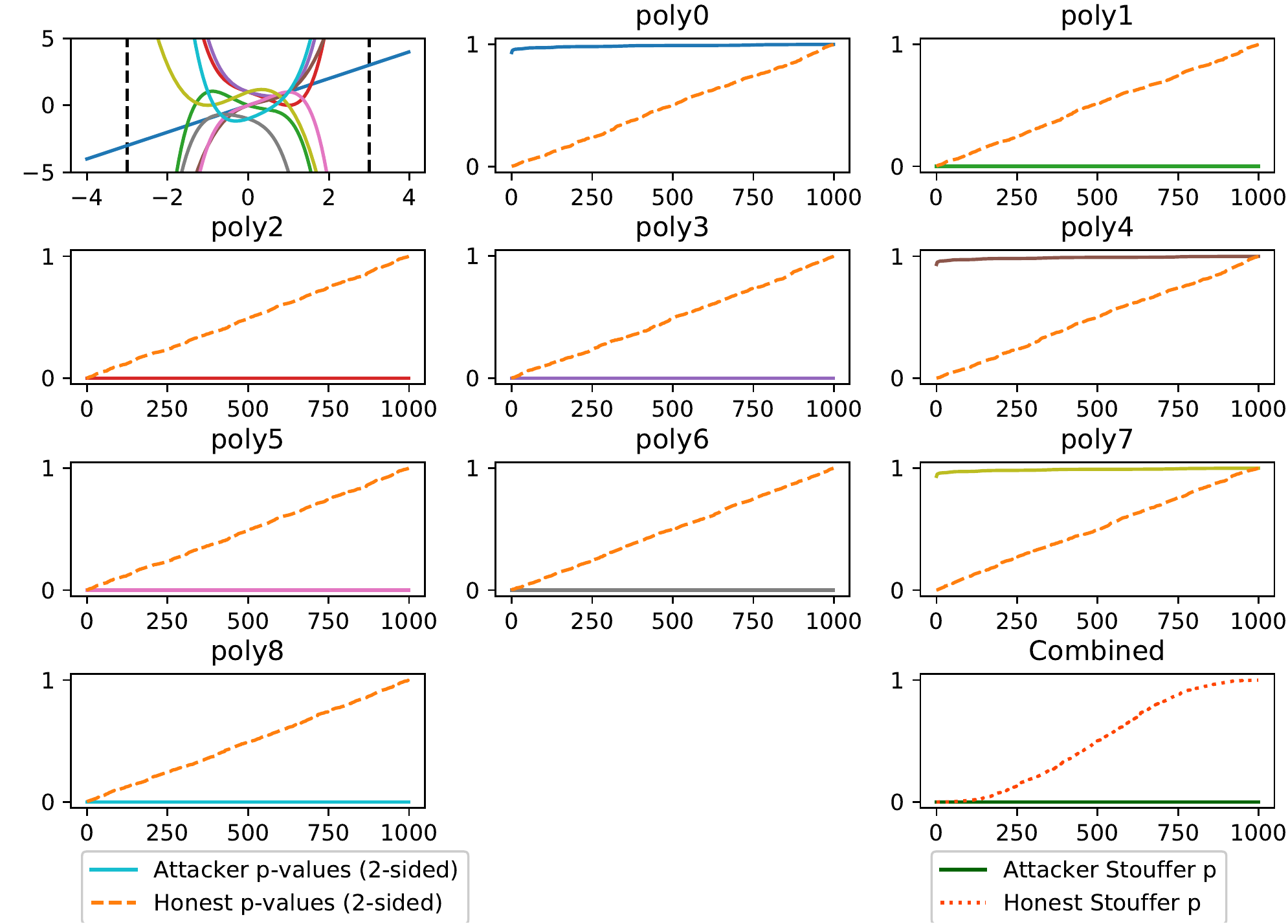}%
\caption{Attack with $q=3$. Defense with polynomials of degree 4.} \label{gamma3}
\end{figure}

\begin{figure}
\includegraphics[width=0.9\textwidth]{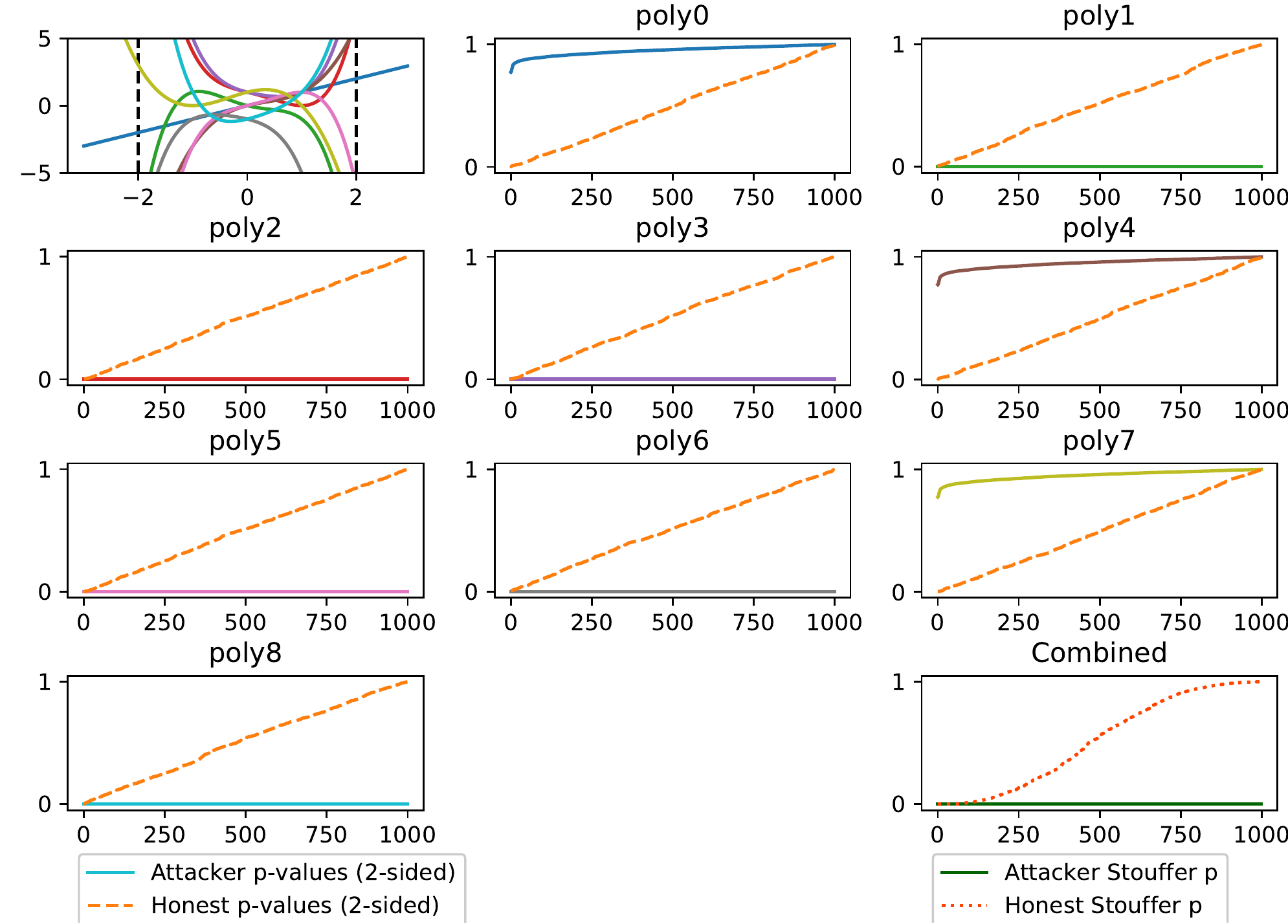}
\caption{Attack with $q=2$. Defense with polynomials of degree 4.} \label{gamma2}
\end{figure}

\begin{figure}
\includegraphics[width=0.9\textwidth]{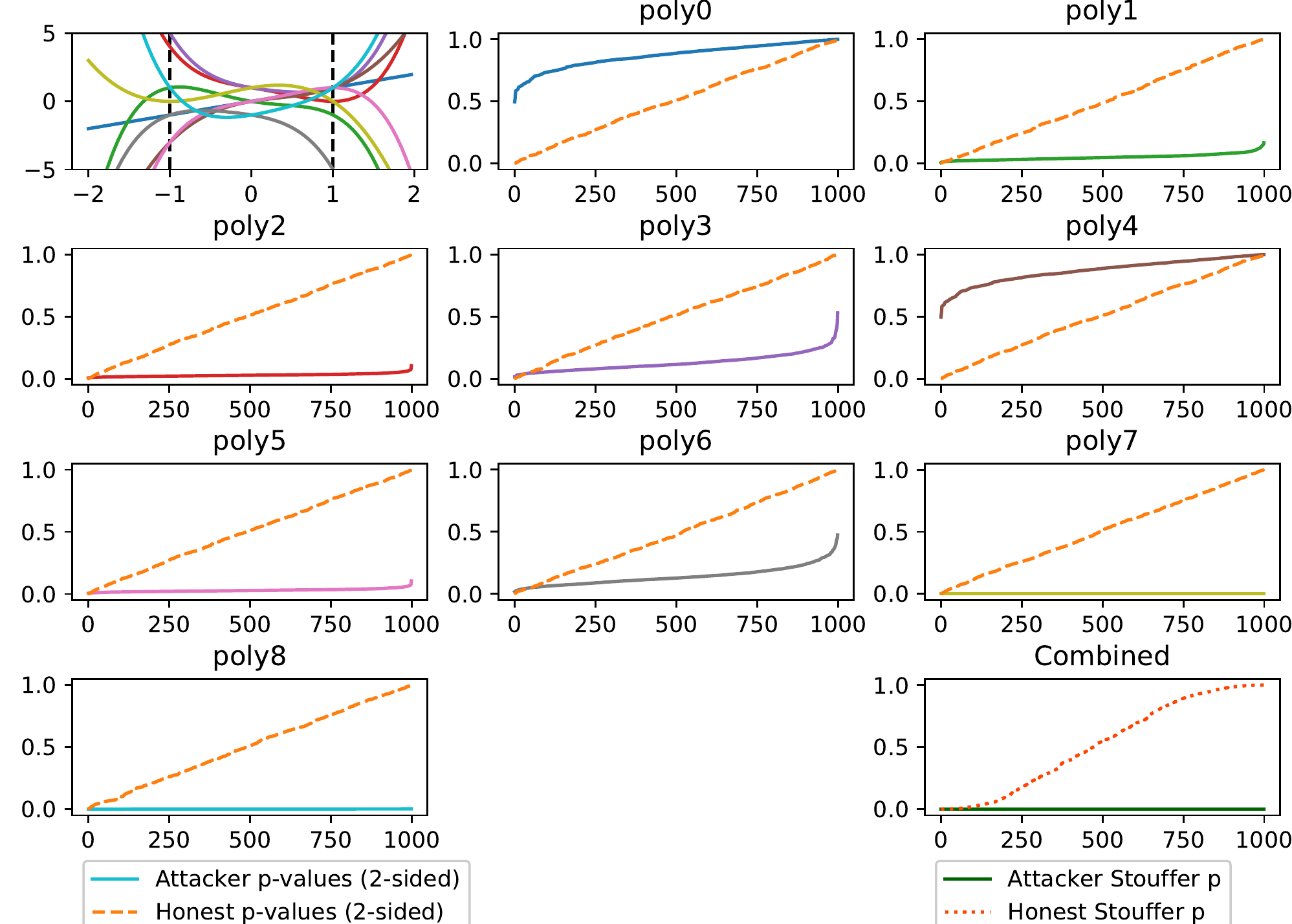}
\caption{Attack with $q=1$. Defense with polynomials of degree 4.} \label{gamma1}
\end{figure}

\begin{figure}
\includegraphics[width=0.9\textwidth]{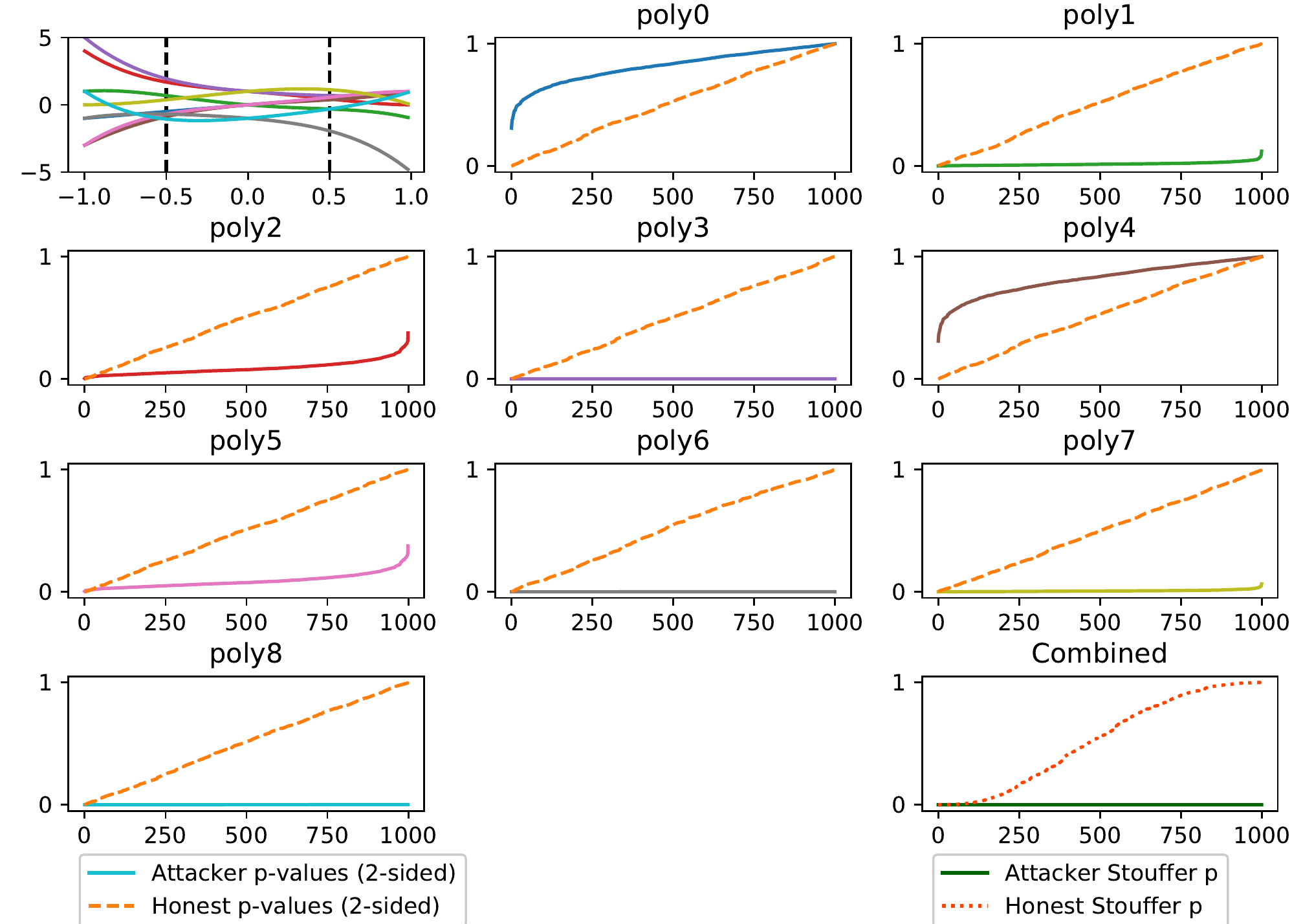}
\caption{Attack with $q=0.5$. Defense with polynomials of degree 4.} \label{gammahalf}
\end{figure}

Experimental results are summarized in Figures \ref{gamma3},~\ref{gamma2},~\ref{gamma1} and~\ref{gammahalf}. \smallskip

The first quadrant illustrates the polynomials used in the simulation and two bars for $\{-q, q\}$. The same random polynomials were used for each experiment. For simplicity, the coefficients of the polynomials were uniformly selected from $\{-1, 0, 1\}$, and (useless)  polynomials of degree lower than 2 were excluded from the random selection. Then, we also added the identity polynomial (\texttt{poly0}), as a witness of what happens when there is no protection.\smallskip

The following nine quadrants give the distribution of $p$-values for each polynomial, over 1000 simulations, sorted in increasing order. The dotted distribution corresponds to what an honest user would obtain, whereas the plain line simulation is based on poisoned datasets.\smallskip

The last quadrant contains the sorted distribution of the aggregated $p$-values using Stouffer. Experimental results show that for poisoned datasets, the aggregated $p$-values remain close to zero, while a honest dataset does not appear to be significantly affected. In other words, with very high probability, keyed testing detects poisoning. \smallskip

\subsection{Discussion}
We observe a saturation when $q$ is too far from $\mu_A$, this is due to the fact that even after passing through $R$ the attack samples remain at the extremes. Hence if $R$ is of odd degree, nothing changes. If the degree of $R$ is even then the two extremes are projected to the same side and Mann-Whitney detects 100\% of the poisonings. It follows that at saturation a keyed Mann-Whitney gives either very high or very low $p$-value. This means that polynomials or B-splines must be carefully chosen to make keying effective.\smallskip

The advantage of combining the $p$-values with Stouffer's method is that the weak $p$-values are very penalizing (by opposition to Pearson's method whose combined $p$-value degrades much slower). A more conservative aggregation would be using Fisher's method.\smallskip

All in all, experimental results reveal that keying managed to endow the test with a significant level of immunity. \smallskip

Interestingly, $\Det$ can be implemented independently of $\mathcal{A}$.\smallskip

A cautionary note: Our scenario assumes that testing does not start before learning is complete. If the opponent can alternate learning and testing then he may infer that a poisonous sample was taken into account (if $\sigma$ was updated and $\mathcal{A}$'s behavior was modified). This may open the door to attacks on $\mathcal{A}$. 

\section{Notes and Further Research}

This paper opens perspectives beyond the specific poisoning problem. e.g. cryptographers frequently use randomness tests $\mathcal{R}$ to assess the quality of random number generators. In a strong attack model where the attacker knows $\mathcal{R}$ and controls the random source it becomes possible to trick many $\mathcal{R}$s into declaring flagrantly non random data as random. Hence, the authors believe that developing \textsl{keyed randomness tests} $\mathcal{R}_{\kappa}$ is important and useful as such.\smallskip 

For instance, in the original minimum distance test 8000 points (a set $S$) sampled from the tested randomness source $\mathcal{S}$ are placed in a $10000\times 10000$ square. Let $\delta$ be the minimum distance between the pairs. If $\mathcal{S}$ is random then $\delta^2$ is exponentially distributed with mean $0.995$. To key the test a secret permutation $R_{\kappa}$ of the plan can generated and the test can be applied to $R_{\kappa}(S)$.\smallskip

To the best of our knowledge such primitives were not proposed yet.\smallskip

We note, however, that keyed protections to different (non cryptographic!) decision problems in very diverse areas do emerge independently e.g. \cite{NaorY14,Kenny,Marius,TRV18}.

\bibliographystyle{splncs04}
\bibliography{bibli}
\newpage

\appendix
\section{Mann-Whitney's $U$-Test}
Let $\mathfrak{D}$ be an arbitrary distribution.\smallskip

Mann-Whitney's $U$-test is a non-parametric hypothesis test. The test assumes that the two compared sample sets $X_0,X_1$ are independent and that a total order exists on their elements (which is the case for real-valued data such as ML feature vectors).\smallskip

Assuming that $X_0\in_R\mathfrak{D}$:

\begin{itemize}
    \item The null hypothesis $H_0$ is that $X_1\in_R\mathfrak{D}$.
    \item The alternative hypothesis $H_1$ is that $X_1\in_R\mathfrak{D'}$ for $\mathfrak{D}\neq\mathfrak{D'}$.
\end{itemize}

The test is consistent\footnote{i.e., its power increases with $\#X_0$ and $\#X_1$.} when, under $H_1$, $P(X_0>X_1) \neq P(X_1>X_0) $.\smallskip

The test computes a statistic called $U$, which distribution under $H_0$ is known. When $\#X_0$ and $\#X_1$ are large enough, the distribution of $U$ under $H_0$ is approximated by a normal distribution of known mean and variance.\smallskip

$U$ is computed as follows:

\begin{enumerate}
    \item Merge the elements of $X_0$ and $X_1$. Sort the resulting list by ascending order.
    \item Assign a rank to each element of the merged list. Equal elements get as rank the midpoint of their adjusted rankings\footnote{e.g., in the list $1, 4, 4, 4, 4, 6$, the fours all get the rank $3.5$}.
    \item Sum the ranks for each set. Let $R_i$ be this sum for $X_i$. Note that if $n_i=\#X_i$ then $R_{1-i}=n(n+1)/2 - R_i$, with $n = n_0 + n_1$.
    \item Let $U_i= R_i - \frac{n_i(n_i+1)}{2}$ and $U=\min(U_0,U_1)$.
\end{enumerate}

When the $\#X_i$ are large enough ($>20$ elements) $U$ approximately follows a normal distribution.\smallskip

Hence, one can check if the value $$z = \frac{U - \lambda_U}{\sigma_U}$$ follows a standard normal distribution under $H_0$, with $\lambda_U$ being the mean of $U$, and $\sigma_U$ its standard deviation under $H_0$: 

$$\lambda_U = \frac{n_0n_1}{2}\mbox{~~and~~}\sigma_U =\sqrt{\frac{n_0n_1(n+1)}{12}}$$

However, the previous formulae are only valid when there are no tied ranks. For tied ranks, the following formula is to be used: 
$$\sigma_{U} = \sqrt{\frac{n_0n_1}{12}\left((n+1)-\sum_{i=1}^k\frac{t_i^3-t_i}{n(n-1)}\right)}$$

Because under $H_0$, $z$ follows a normal distribution, we can estimate the likelihood that the observed values comes from a standard normal distribution, hence getting a related $p$-value from the standard normal table. \smallskip

\section{Stouffer's $p$-Value Aggregation Method}
$p$ values can be aggregated in different ways \cite{heard2018choosing}. Stouffer \cite{SSD49} observes that the $z$-value defined by $z=\Phi^{-1}(p)$ is a standard normal variable under $H_0$ where $\Phi$ is the standard normal CDF. Hence when $\{p_1,\dots,p_{\ell}\}$ are translated into $\{z_1,\dots,z_{\ell}\}$, we get a collection of independent and identically distributed standard normal variables under $H_0$. To combine the effect of all tests we sum all the $z_i$ which follows a normal distribution under $H_0$ with mean $0$ and variance $\ell$. The global test statistic
\[
Z = \frac{1}{\ell}\sum_{i=1}^{\ell} z(p_i)
\]
is hence standard normal under $H_0$ and can thus be reconverted into a $p$-value in the standard normal table. \smallskip

Note that in theory, combining $p$-values using Stouffer's method requires that the tests are independent. Other methods can be used for combining $p$-values from non-independent tests, e.g. \cite{KOST2002183,10.2307/2529826}. However, these calculations imply that the underlying joint distribution is known, and the derivation of the combination statistics percentiles requires a numerical approximation.

\end{document}